\documentclass[sigconf]{acmart}
\AtBeginDocument{%
  \providecommand\BibTeX{{%
    \normalfont B\kern-0.5em{\scshape i\kern-0.25em b}\kern-0.8em\TeX}}}

\setcopyright{rightsretained}
\copyrightyear{2021}
\acmYear{2021}
\acmDOI{}
\acmISBN{} 

\begin{document}

\title{\textit{Mobilkit}: A Python Toolkit for Urban Resilience and Disaster Risk Management Analytics using High Frequency Human Mobility Data}

\author{Enrico Ubaldi}
\email{enrico.ubaldi@mindearth.org}
\affiliation{%
  \institution{MindEarth}
  \country{}
}

\author{Takahiro Yabe}
\email{tyabe@purdue.edu}
\affiliation{%
  \institution{Purdue University, World Bank}
  \country{}
}

\author{Nicholas K. W. Jones}
\email{njones@worldbankgroup.org}
\affiliation{%
  \institution{Global Facility for Disaster Reduction and Recovery (GFDRR) / World Bank}
  \country{}
}

\author{Maham Faisal Khan}
\email{mkhan57@worldbank.org}
\affiliation{%
  \institution{GFDRR / World Bank}
  \country{}
}

\author{Satish V. Ukkusuri}
\email{sukkusur@purdue.edu}
\affiliation{%
  \institution{Purdue University, World Bank}
  \country{}
}

\author{Riccardo {Di Clemente}}
\email{r.di-clemente@exeter.ac.uk}
\affiliation{%
  \institution{Department of Computer Science, University of Exeter}
  \country{}
}

\author{Emanuele Strano}
\email{emanuele.strano@mindearth.org}
\affiliation{%
  \institution{MindEarth}
  \country{}
}

\renewcommand{\shortauthors}{Ubaldi, et al.}

\begin{abstract}
Increasingly available high frequency location datasets derived from smartphones provide unprecedented insight into trajectories of human mobility. These datasets can play a significant and growing role for informing preparedness and response to natural disasters. However, limited tools exist to enable rapid analytics using mobility data, and tend not to be tailored specifically for disaster risk management. We present an open-source, Python-based toolkit designed to conduct replicable and scalable post-disaster analytics using GPS location data. Privacy, system capabilities, and potential expansions of \textit{Mobilkit} are discussed.
\end{abstract}



\maketitle

\section{Introduction}

 Data on human mobility patterns derived from mobile phones has played a significant and growing role in informing preparedness and response to natural disasters. Early applications \cite{gonzalez2008understanding,blondel2015survey,iqbal2014development} - such as responses to the Haiti and Nepal earthquakes (2010 and 2015 respectively) made use of Call Detail Records (CDR): a form of data recorded by mobile network operators for billing purposes. However, analysts face barriers to using CDR data related to access, as well as limited spatial and temporal resolution dictated by spacing of cellphone towers and the frequency with which users place calls. More recently, a newer generation of smartphone-derived human mobility data has become available. Privacy-protected data products derived from smartphone apps provide information on human movement patterns at higher spatial and temporal resolution and with less onerous data access challenges compared to CDR, unlocking several scalable use cases to inform disaster risk management (DRM) \cite{yabe2020quantifying,lu2012predictability,wilson2016rapid}. This data, too, can prove challenging to use. The data tends to be `big' (typically 50 GB for just one month of data for a few thousands users); and while several open-source tools exist \cite{pappalardo2019scikitmobility,graser2019movingpandas,de2016bandicoot} to conduct analysis with human mobility datasets, none provide specialized functionality needed to address key questions concerning natural disaster impacts on urban areas and most of them did not provide a parallel, high-performance implementation. Some providers of mobility data pre-aggregate the data to circumvent barriers to entry, e.g. Facebook Disaster Maps\footnote{\url{https://dataforgood.fb.com/tools/disaster-maps/}}, but these come at the cost of the flexibility of disaggregated data. In this work, we present a toolkit designed to conduct replicable and scalable analyses for DRM and resilient urban planning utilizing large human mobility datasets in the form of raw (disaggregated) GPS pings.
 
 \begin{figure*}[h]
  \centering
  \includegraphics[width=\textwidth]{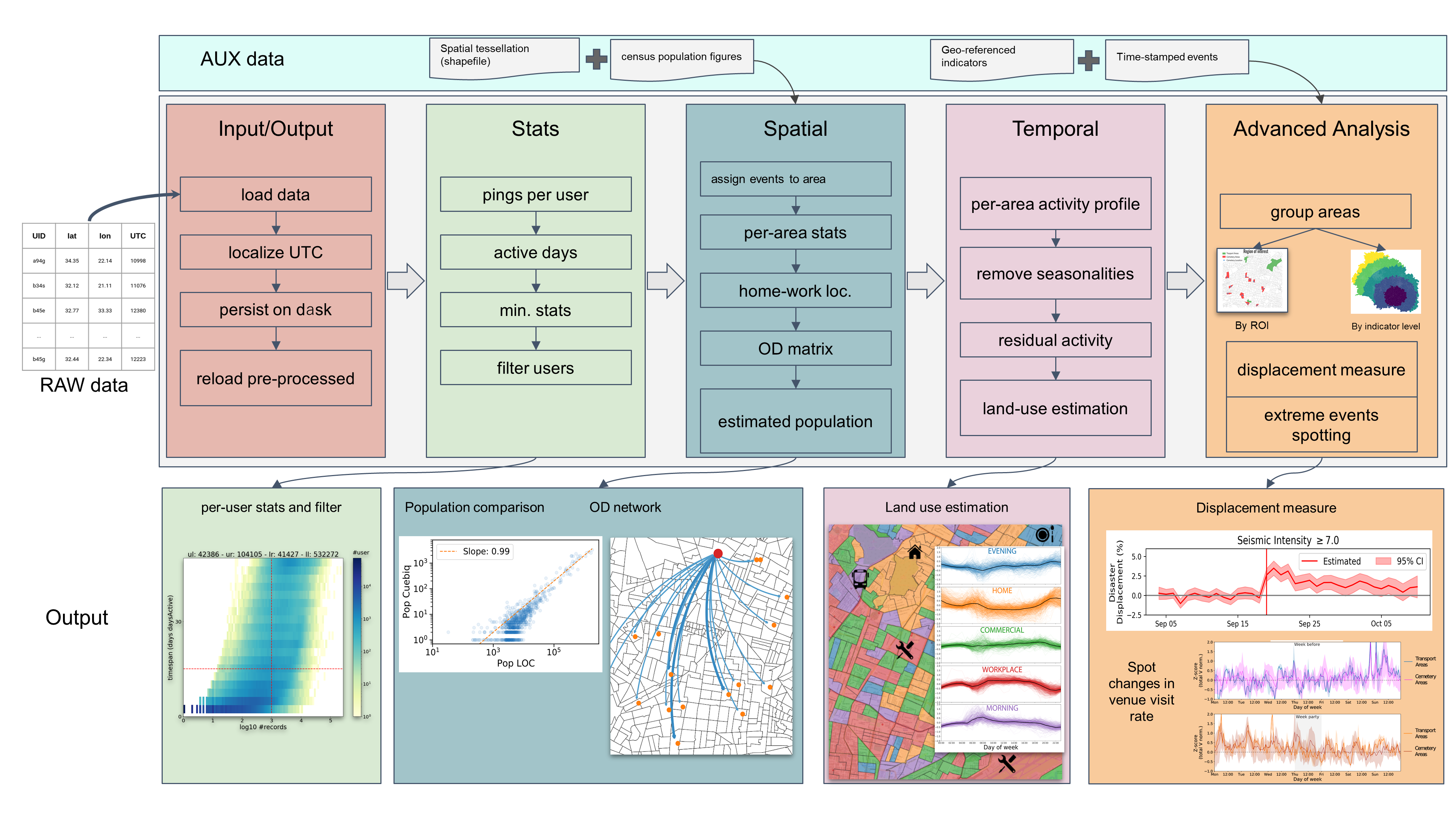}
  \caption{Analytical capabilities of \textit{Mobilkit}}. 
  \label{mobilkit}
\end{figure*}
 
 \section{Solution}

\subsection{Analytical Capabilities.} \textit{Mobilkit} provides an end-to-end computational framework---depicted schematically in Figure \ref{mobilkit}---to use location data for DRM analytics, from loading, pre-processing, descriptive statistics and analysis, up to visualization and output metrics.The toolkit enables users to (1) load and filter raw mobility data covering large spatial and temporal extensions, (2) compute user statistics (number of active days, number of positions recorded) and filter them according to the required analysis, (3) extract home and work locations of users based on a given tessellation, using mean-shift clustering algorithms, (4) compute the land use of a given urban region using hierarchical clustering techniques, and (5) characterize the displacement of people under different groupings (e.g. distance from an epicenter, socio-economic index, etc.) after a major disaster event.
\textit{Mobilkit} enables efficient computation of population displacement after extreme events (2h to compute displacement rates of 130,000 unique users over 5 weeks of data, using a laptop with 12 cores and 32GB memory). By overlaying other context layers, the library also allows rapid analyses of how spatial and socio-economic characteristics affect displacement. Moreover, the tool facilitates mapping density of home and work locations; creating origin-destination matrices of commuting patterns; and tracking changes in visit rates to relevant Points-of-Interest (POIs). In an upcoming paper, \cite{mexicopaper}, we illustrate a case study that makes use of \textit{Mobilkit} to understand and quantify the impact of the 2017 Puebla Earthquake in Mexico.

\subsection{System Design.} 
\textit{Mobilkit} is written in Python, owing to the language’s wide user base and large community of open-source contributors \cite{srinath2017python}. The library builds on the Dask framework~\cite{rocklin2015dask} to facilitate computation of `big’ mobility datasets on limited compute resources through parallel batch computing. Indeed, Dask (as other high performance data analysis frameworks such as Spark) organizes the computation in a graph of tasks that get applied to small batches of data, enabling the analysis of datasets larger than memory also on personal computers. Moreover, the Dask framework allows to define user defined functions (UDF) to aggregate grouped data or reduce partitions of a dataset using standard Python functions and modules. This provides a simpler interface with respect to more complex frameworks such as pySpark, that requires the user to either learn a new language to implement a custom UDF or to use a verbose, hard to debug interface between Python and the native Spark code.
The Dask framework also allows to speed up most of the common analyses required: for example, on a sample of 230'000 users and 57 millions pings the time needed to compute the basic users' statistics drops from $\sim 900$ seconds with a single core to 200 seconds when using 8 parallel threads (same figures holds when computing the activity profiles of single areas).
Of course, Dask comes with some limitations. For instance, it is a less mature, partially undocumented framework with respect to Spark, which is an industry standard. Moreover Dask does not offer a complete SQL-like interface and a large array of extensions to implement ad-hoc analysis as Spark does (such as with the spatial data analysis library  Sedona\footnote{https://sedona.apache.org/}), and its API are limited to Python. Nevertheless, the ease of use and the possibility to easily implement new functions let us choose Dask over other implementations.
The code is open-source,  released under the MIT license and can be accessed on Github (\url{https://github.com/mindearth/mobilkit}), with complete documentation, examples and tutorials in the form of jupyter notebooks (\url{https://mobilkit.readthedocs.io/en/latest/}). Mobilkit can be easily installed on the command line, with \texttt{pip install mobilkit}.

\section{Discussion}
Existing open-source libraries provide a suite of analytical tools to analyze GPS trajectories, including radius of gyration, entropy of places, social networks (e.g. clustering coefficient). However, these are either tailored to more traditional, CDR data \cite{de2016bandicoot}, or are not specialized for post-disaster analytics \cite{pappalardo2019scikitmobility,graser2019movingpandas}. \textit{Mobilkit} provides additional functionality to compute displacement of populations after extreme events such as floods and earthquakes, identify anomalies in population density, and infer land use before and after a disaster. 

Users of these datasets must remain aware of the privacy-preservation challenges inherent in disaggregated location data. A record of user-level data typically comprises GPS coordinates (sometimes accompanied by an accuracy measure in meters or degrees), along with a timestamp and anonymized user ID. However, established methodologies \cite{Xu_2017} argue that, given the uniqueness of user-level trajectories, user ID-level anonymizations are possible to circumvent. Whilst \textit{Mobilkit} provides aggregations that add a simple layer of privacy over this, these aggregations alone may not meet privacy protections according to typical regulatory or IRB standards. Further care is needed for users of the toolkit to ensure sensitive information is not divulged during analysis.

Moving forward, we plan to expand the functional capabilities of \textit{Mobilkit} for analyzing urban accessibility, urban segregation, and transportation network resilience. We seek also to diversify the compatible data types to include flow-based mobility data, trips and stops sequences and spatially-aggregated mobility data.

\begin{acks}
This work has been conducted under a grant from the Spanish Fund for Latin America and the Caribbean (SFLAC) under the Disruptive Technologies for Development (DT4D) initiative. We extend our thanks to Cuebiq Inc which provided privacy-protected datasets used while developing the library. The findings, interpretations, and conclusions expressed in this paper are entirely those of the authors. They do not necessarily represent the views of the International Bank for Reconstruction and Development/World Bank and its affiliated organizations, or those of the Executive Directors of the World Bank or the governments they represent. 
\end{acks}

\bibliographystyle{ACM-Reference-Format}
\bibliography{main}

\end{document}